\documentclass[a4paper,10pt,twoside]{cpc-hepnp}
\usepackage{epstopdf}
\usepackage{multicol}
\usepackage{graphicx}
\usepackage[colorlinks,linkcolor=blue,anchorcolor=blue,citecolor=blue,hyperindex,CJKbookmarks,dvipdfm]{hyperref}
\usepackage{booktabs}
\usepackage{amssymb,bm,mathrsfs,bbm,amscd}
\usepackage[tbtags]{amsmath}
\usepackage{lastpage}
\usepackage{CJK}

\begin{document}
\begin{CJK*}{GBK}{song}
\bibliographystyle{unsrt}

\fancyhead[c]{\small Chinese Physics C} \fancyfoot[C]{\small \thepage}

%\footnotetext[0]{Received 14 March 2009}

\title{Transverse beam size measurement system using visible synchrotron radiation at HLS II\thanks{Supported by National Natural Science Foundation of China (11105141, 11175173) and the upgrade project of Hefei Light Source}}

\author{%
TANG Kai()1$^{1;1)}$\email{tangkkai@mail.ustc.edu.cn}%
\quad SUN Bao-Gen()$^{1;2)}$\email{bgsun@mail.ustc.edu.cn}%
\quad YANG Yong-Liang()$^{1}$
\quad LU Ping()$^{1}$\\
\quad TANG Lei-Lei()$^{1}$
\quad WU Fang-Fang()$^{1}$
\quad CHENG Chao-Cai()$^{1}$\\
\quad ZHENG Jia-Jun()$^{1}$
\quad LI Hao()$^{1}$
}
\maketitle

\address{%
$^1$ National Synchrotron Radiation Laboratory, University of Science and Technology of China, Hefei 230029, China\\
}

\begin{abstract}
 An interferometer system and an imaging system using visible synchrotron radiation (SR) have been installed in HLS II storage ring. Simulations of these two systems are given using Synchrotron Radiation Workshop(SRW) code. With these two systems, the beam energy spread and the beam emittance can be measured. A detailed description of these two systems and the measurement method is given in this paper. The measurement results of beam size, emittance and energy spread are given at the end.
\end{abstract}



\footnotetext[0]{\hspace*{-3mm}\raisebox{0.3ex}{$\scriptstyle\copyright$}2013
Chinese Physical Society and the Institute of High Energy Physics
of the Chinese Academy of Sciences and the Institute
of Modern Physics of the Chinese Academy of Sciences and IOP Publishing Ltd}%

\begin{multicols}{2}

\section{Introduction} \label{Introduction}

Visible synchrotron radiation (SR) has been widely used for electron beam diagnostics in the storage rings. Overview of transverse beam profile diagnostic based emitted SR are given by Kube \cite{1} and Takano \cite{2} respectively. Beam diagnostic methods based SR can be classified in imaging methods, exploitation of SR wave-optics feature methods(such as $\pi$-polarization), projection methods and interference methods \cite{3}. The measurement resolution is limited in a simple imaging system(using visual SR) due to diffraction effects \cite{1}. However, a simple imaging system was still applied at the B8 beamline of HLS II \cite{4} because the imaging system can monitor the beam state directly. Besides, HLS II is a second generation SR source with several hundreds $\mu$m transverse beam size both horizontal and vertical. Error results by those effect can be no more than 5$\%$.

The SR interferometer is first applied to measure the beam size by Mitsuhashi at the ATF damping ring \cite{5,6,7,8}. Now it has become a universal tool and has been applied in numerous facilities \cite{9,10,11,12,13,14,15,16,17,18}. It was successfully demonstrated to measure a beam size of $39.0 \mu m(H) \times 14.7 \mu m(V)$ with a resolution better than 1 $\mu$m \cite{7}. T. Naito and T. Mitsuhashi \cite{19} apply an interferometer with Herschelian reflective to reduce the disperse effect of the objective lens. By this method, the interferometer can measure a 4.7 $\mu$m beam size. P. Chevtsov utilizes an interferometer to measure the beam energy spread by ignoring the intrinsic beam size \cite{9}. At HLS II, an interferometer system has been installed at the B7 beamline and a simple imaging system at the B8 beamline\cite{4,20}.

With knowledge of the machine optical parameters and relative energy spread, the emittance can be inferred from the measured beam size \cite{1}. Usually, the beam energy is not a constant and depends on beam current. So emittance is not a direct value. In this paper, a nondestructive method to measure beam energy spread and emittance simultaneously using these two systems is presented.

The configuration of these two systems are described in Section \ref{Experimental setup}. The measurement method of the beam energy spread and the horizontal beam emittance using these two systems is described in Section \ref{Measurement theory}. The simulation results computed by SRW \cite{21} is given in \ref{Simulation Result}. The measured result is given in Section \ref{Results and analysis}.

\section{Experimental methods} \label{Experimental methods}
\subsection{Experimental setup} \label{Experimental setup}
HLS II is a second generation electron storage ring with an electron energy of 800 MeV. It has a fourfold periodicity with total eight 45$^{\circ}$ sector magnets, a circumference of 66 m, a design transverse beam emittance of 36.4 nm.rad and a design relative energy spread of 4.7 $\times10^{-4}$. Tab.\ref{tab1} lists the beam parameters and the theoretic beam sizes at the B7 and B8 source point.

\begin{center}
\tabcaption{ \label{tab1}  Optics parameters and beam size of the B7 and B8 source point.}
%\footnotesize
\begin{tabular*}{80mm}{c@{\extracolsep{\fill}}ccc}
%$\toprule Parameters & $\beta_{x}$(m) & $\beta_{y}$(m) & $\alpha_{x}$ & $\alpha_{y}$ & $\eta_{x}$ & $\eta_{x}$ \\
%\toprule Mass & $\sigma$/mb   & $\rho$  & \% Error \\
\toprule Parameters & B7   & B8  \\
\hline
$\beta_{x}$(m) & 1.7668  & 0.9235 \\
$\beta_{y}$(m) & 12.3485 & 7.6854 \\
%$\alpha_{x}$   & -3.002  & 1.3788\\
%$\alpha_{y}$   & 2.1319 & -2.0098\\
$\eta_{x}$(m)  & 0.1059 & 0.2793\\
$\eta'_{x}$    & -0.1990 & -0.4754\\
Design transverse $\varepsilon$(nm.rad) & 36.4 &\\
Design energy spread $\delta$ & $4.7 \times 10^{-4}$ \\
$\sigma_{x}$(mm) with 5$\%$ coupling & 252.4 & 221.9 \\
$\sigma_{y}$(mm) with 5$\%$ coupling & 146.3 & 115.4 \\
$\sigma_{x}$(mm) with 10$\%$ coupling & 246.9 & 218.6 \\
$\sigma_{y}$(mm) with 10$\%$ coupling & 202.1 & 159.5 \\
\bottomrule
\end{tabular*}
\end{center}

Fig.~\ref{fig1} shows the schematic layouts of the interferometer system. The SR is reflected 90$^\circ$ downwards by an oxygen-free copper mirror(mirror1) in the vacuum chamber. %The diameter of mirror1 has a diameter of 90 $mm$ and the angular acceptance of this area is 8.45 mrad $\times$ 5.98 mrad(\emph{H}$\times$\emph{V}).
After reflected by Mirror2 and mirror3, the SR is transmitted to an optics table. On the optical table, the SR light is divided by the splitter1. The SR of these two channels passes through a double slit, an achromatic lens with 1000 mm focal length, a lens2 with 100 mm focal length, a polarizer and a 500 nm bandpass filter respective. The image is observed by a Procilica GE680 camera which is placed at the image plane.

The distance between the B7 source point and the double slit are 10.8 m for the horizontal channel and 11.1 m for the vertical channel. The slit separation is 5 mm and the size of the aperture is 0.5 mm $\times$ 2 mm(\emph{H$\times$V}) for the horizontal channel and 2 mm $\times$ 0.5 mm(\emph{H$\times$V}) for the vertical channel. With adjusting the place of Lens2 and the camera, we can magnify the pattern image by 3$\thicksim$5 ratios. The bandpass filter with a central wavelength at 500 nm and 10 nm FWHM bandwidth.

%Mirror1 is water-cooled so that it does not suffer thermal deformations form the SR heat load. The surface of mirror1 is polished and the 70$\%$ area at the center has a flatness specification of 0.22 $\lambda$ at 500 nm so that it satisfies Rayleigh's criterion. The polarizer can be rotated manually to choose either the $\sigma$-polarized or $\pi$-polarized component of SR.
Fig.~\ref{fig2} shows the schematic layouts of the imaging system. Mirror1 is a water-cooled oxygen-free copper. Mirror2 $\sim$ Mirror5 are reflect mirror. The achromatic lens has a focal length of 1400 mm. After reflected by a periscope layout(composed by mirror4 and mirror5), the SR is transmitted to an optics table. The lens2 is used to adjust the magnification of the imaging system.

The focal length of all lenses were measured. Local bump experiment was performed to measure the magnification of those two systems. The double slit was removed  at the experiment. Experiments of dispersion function measurement were achieved by changing RF. Betatron functions used are theoretical value.

The camera has a CCD of 480 pixels $\times$ 640 pixels with pixel size of 7.4 ${\mu}m$ $\times$ 7.4 ${\mu}m$. Experiment was performed to make sure that output property of the CCD is linear. Experiment were also performed by analysing the image obtained by the camera at different exposure time. The results show that the beam size measurement result doesn't depend on the exposure time of the camera.

\begin{center}
\includegraphics[width=8cm]{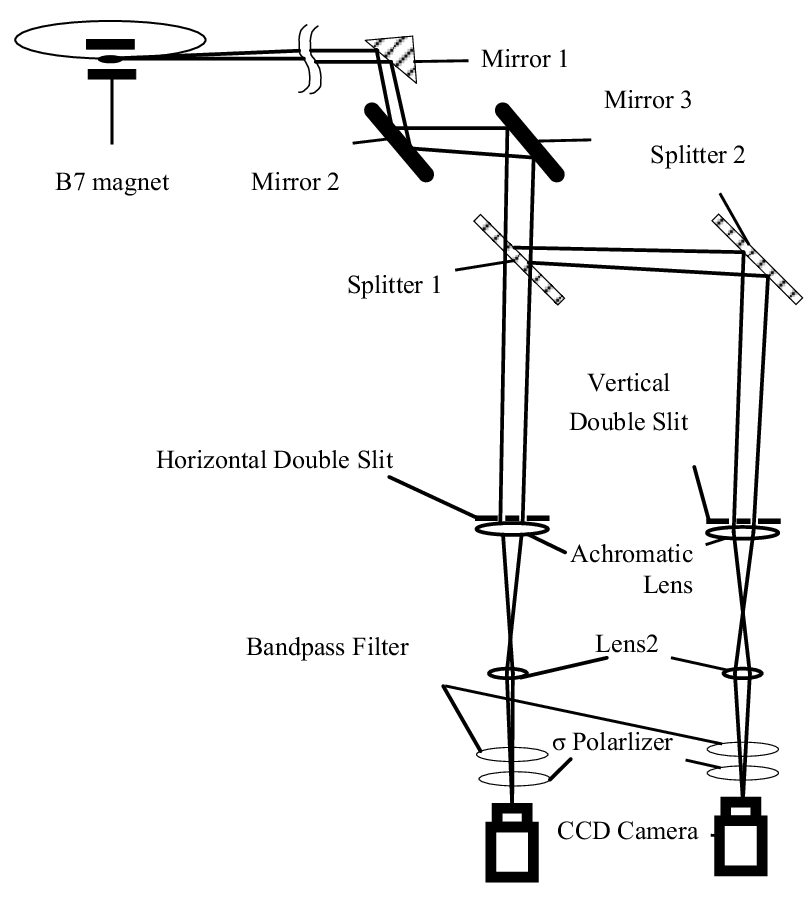}
\figcaption{\label{fig1}   Layout of the interferometer system at the B7 beamline}
\end{center}

\begin{center}
\includegraphics[width=8cm]{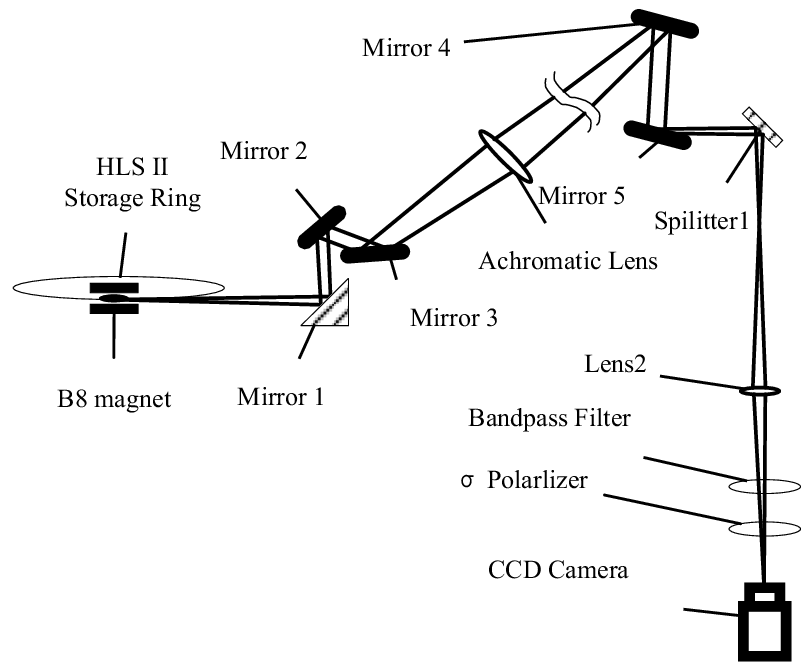}
\figcaption{\label{fig2}   Layout of the imaging system at the B8 beamline}
\end{center}

\subsection{Measurement theory} \label{Measurement theory}
The theory of beam size measurement with interferometer is well-documented. It is a wavefront division two-beam interferometer using polarized quasi-monochromatic light. The SR wavefront is divided by a double slit and merge again at the image plane. Take the vertical channel as example, the fringe distribution \cite{7} at the image plane is described by:
\begin{equation}\label{eq1}
    I(y)=I_{0}{{\rm{sinc}}^2}({\frac{{\pi{w_y}}}{\lambda{L^{'}}}}y)(1+{\gamma_{y}}\cos(\frac{2\pi{d_{y}}}{\lambda{L^{'}}}y+\phi_{0}))
\end{equation}
Where $w_y$ denotes vertical size of the slit, $L^{'}$ denotes the distance between the double slits and the detector, $d_{y}$ denotes the separation of the two slits, $\lambda$ denotes the wavelength, $\phi_{0}$ denotes the fringe phase, and $\gamma_{y}$ denotes the complex degree of coherence. If the beam shape is a Gaussian profile \cite{11}, $\gamma_{y}$ is given by:
\begin{equation}\label{eq2}
    \gamma_{y}=\exp\{-2({\frac{d_y \sigma_y}{\lambda L}})^2\}
\end{equation}
Where $\sigma_y$ denotes the beam size. Expanding Eq.~\eqref{eq2}:
\begin{equation}\label{eq3}
    \sigma_y =  \frac{\lambda L}{\pi d_y} \sqrt{\frac{1}{2} \ln \frac{1}{\gamma_y}}
\end{equation}
Reality, the fringe was fitted by function model below:
\begin{equation}\label{eq4}
    I(y)= a_1 {\rm sinc}^2 (a_2 (y- a_5 )) \{1+ a_3 \cos[a_4 (y-a_6 )] \}+a_7
\end{equation}
where sinc denotes $\sin(x)/x$ function, $a_1, a_2, ..., a_7$ are fitting parameters. And $a_3$ is $\gamma_{y}$.

The theory of the transverse beam size measurement with the imaging system is simple. Ideally, the intensity profile monitored by camera would correspond to the beam profile at the source point scaled by the magnification factor of the system \cite{15}. The vertical beam size is 115.4 ${\mu}m$ and horizontal beam size is 221.9 ${\mu}m$ (Table.\ref{tab1}) when the coupling coefficient is 5$\%$. Thus, the resolution limited by diffraction is far smaller than the beam size. With integrating the raw data along x direction and y direction respectively, two curves can be achieved. The beam size can be directly inferred by fitting these curves with function:
\begin{equation}\label{eq5}
    I(x)= a_{1} \exp(- \frac{{(x-a_{2})}^2}{2{a_{3}}^2})+a_{4}
\end{equation}
where $a_{1}$ is related to light intensity, $a_{2}$ is the peak position of the curve and related to beam position, $a_{3}$ is the beam size, $a_{4}$ is related to the camera noise. The beam size is inferred by dividing $a_{3}$ by magnification ratio. This fitting model takes account of the camera noise and the background from stray illumination. An offset must be obtained if one fitted the profile with the Gaussian function.

Usually, the horizontal beam size has two sources: the portion due to betatron oscillation and the portion due to dispersion. Let $\varepsilon_x$ denote horizontal beam emittance and $\delta$ denotes relative energy spread.
 \begin{equation}\label{eq6}
   \varepsilon_x = \frac{\sigma_x^2 - \delta^2 \eta_x^2 }{\beta_x}
 \end{equation}
Where $\beta$ is betatron function and $\eta$ is dispersion function. With subtracted the part due to dispersion, beam emittance can be inferred. Let $\sigma_\beta = \sqrt{\varepsilon_x \beta_x}$ denotes the size due to betatron oscillation and $\sigma_\delta = \delta \eta$ denotes the size due to dispersion. For B8 source point, the $\sigma_\beta$(178.9 $\mu m$) can be compare to the $\sigma_\delta$(131.3 $\mu m$). For the B7 source point, the $\sigma_\delta$(49.8 $\mu m$) is far less than the $\sigma_\beta$( 247.4 $\mu m$).

For the storage ring, the beam emittance and energy spread do not depend on the longitudinal position. Thus, horizontal beam emittance and energy spread at the B7 and B8 source point are the same. They can be described by:
\begin{equation}\label{eq7}
   \left\{ \begin{array}{l}
{\sigma _{x,1}}^2 = {\varepsilon _x}{\beta _{x,1}} + {\delta ^2}{\eta _{x,1}}^2\\
{\sigma _{x,2}}^2 = {\varepsilon _x}{\beta _{x,2}} + {\delta ^2}{\eta _{x,2}}^2
\end{array} \right.
 \end{equation}
Where subscript 1 represent B7 and subscript 2 represent B8. We can solve for $\varepsilon_x$ and $\delta$ from Eq.~\eqref{eq7}:
 \begin{equation}\label{eq8}
\left\{ \begin{array}{l}
{\varepsilon _x} = ({\sigma _{x,1}}^2{\eta _{x,2}}^2 - {\sigma _{x,2}}^2{\eta _{x,2}}^2)/({\beta _{x,1}}{\eta _{x,2}}^2 - {\beta _{x,2}}{\eta _{x,1}}^2)\\
\delta  = [({\sigma _{x,2}}^2{\beta _{x,1}} - {\sigma _{x,1}}^2{\beta _{x,2}}) / ({\beta _{x,1}}{\eta _{x,2}}^2 - {\beta _{x,2}}{\eta _{x,1}}^2)]^{1/2}
\end{array} \right.
 \end{equation}

After the horizontal beam sizes at the B7 and B8 source point measured simultaneously, the corresponding horizontal beam emittance and the beam energy spread can be inferred by Eq.~\eqref{eq8}.
\subsection{Simulation Result} \label{Simulation Result}
The SRW code can readily compute the SR emission \cite{19}. The intensity distribution caused by a single electron is called filament-beam-spread function(FBSF). And the SR intensity distribution is computed by making a 2D convolution of the FBSF with a 2D Gaussian distribution profile.

The theoretical beam size at the B7 source point is 252.4 $\mu m(H) \times 146.3 \mu m(V)$(Tab.~\ref{tab1}) and the corresponding degree of coherence are 0.3586 and 0.7007 respective when the coupling coefficient is 5$\%$. The value of degree value of coherence(Tab.~\ref{tab2}) can be inferred by fitting the simulated fringe with the function described in Eq.~\eqref{eq4}. The simulation $\gamma_x$ is 0.6909 when the $\sigma$-polarized component of the SR is selected to illuminate the double slit and 0.6871 for $\pi$-polarized and 0.6825 for total SR. The simulation $\gamma_y$ hardly depend on the polarity of SR.

\begin{center}
\tabcaption{ \label{tab2}  The degree of coherence simulated by SRW using different polarization component of the SR(theoretic value are 0.3586(H) and 0.7007(V)).}
\begin{tabular*}{80mm}{c@{\extracolsep{\fill}}ccc}
\toprule Parameters        & $\sigma$-polarized   & $\pi$-polarized & total  \\
\hline
$\gamma_x$       &0.3502    &0.3503     &0.3502\\
Relative error($\%$)        &2.3       &2.3        &2.3\\
$\gamma_y$                  &0.6909    &0.6871     &0.6825\\
Relative error($\%$)         &1.4       &1.9        &2.6\\
\bottomrule
\end{tabular*}
\end{center}

The simulated degree of coherence is less than the theoretic one(Tab.~\ref{tab1}). Because the wavefront of SR was sampled at a finite area so that some beam profile information is lost. Experiment of changing the sampling area size was performed. The bigger the sampled area is, the more accurate the simulation result will be.

\begin{center}
\includegraphics[width=8cm]{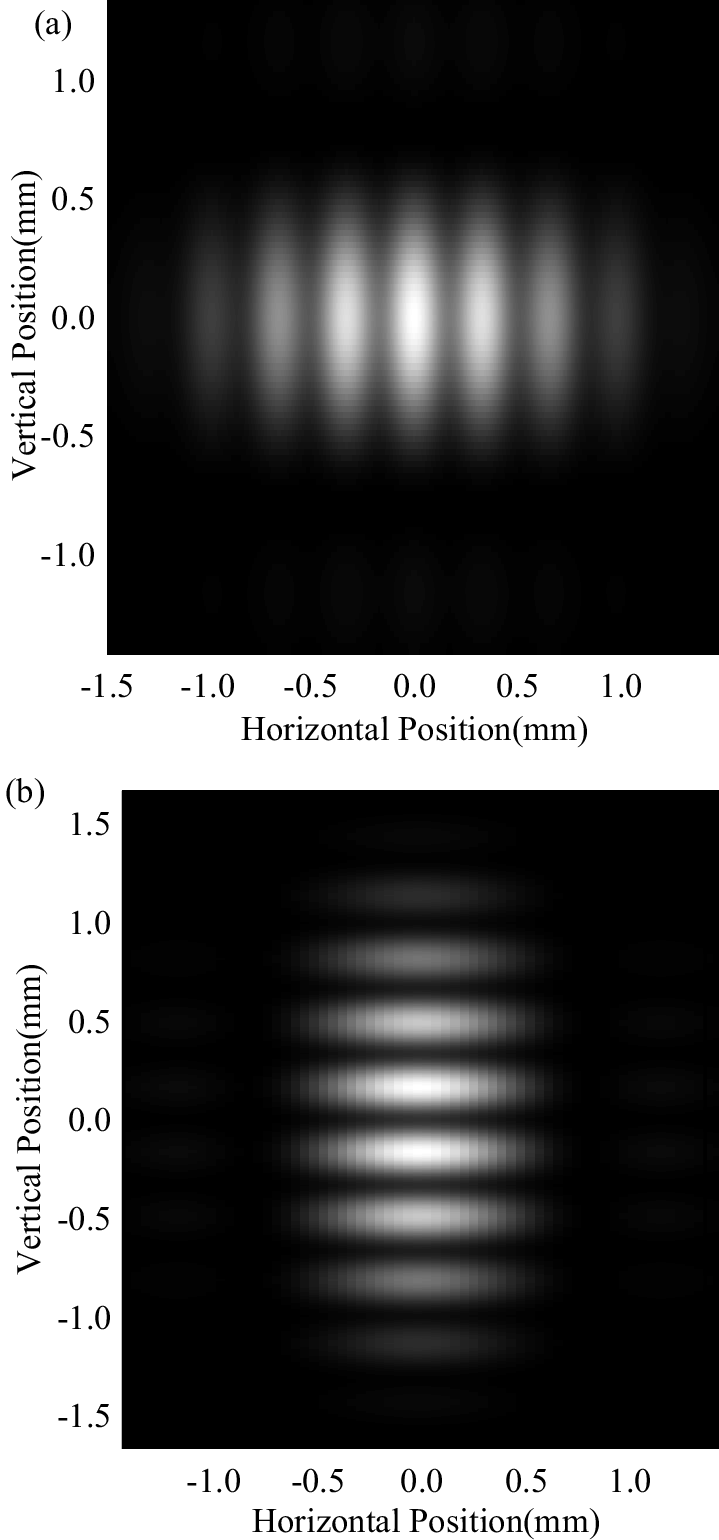}
\figcaption{\label{fig3} The simulated(using SRW) intensity distribution of $\sigma$-polarization SR at the image plane of the interferometer system. (a)Horizontal channel assuming a finite beam ($\sigma_{x} =253.4 \mu m$). (b) Vertical channel assuming a finite beam ($\sigma_{y} =149.9 \mu m$)}
\end{center}

For imaging system, the simulation FBSF is $17.4 \mu m(H) \times 26.8 \mu m(V)$. The simulation beam size is $225.8 \mu m(H) \times 122.2 \mu m(V)$ for $\sigma$-polarization, $225.7 \mu m(H) \times 127.8 \mu m(V)$ for $\pi$-polarization and $225.7 \mu m(H) \times 122.7 \mu m(V)$ for total polarization. The simulation error for vertical direction of the imaging system has two sources: finite size of sampling area and fitting model error. The intensity distribution of imaging with $\pi$-polarized is a curve which has two peaks instead of Gaussian distribution. By comparing the simulation result with the theoretical value ($221.7 \mu m(H) \times 118.3 \mu m(V)$), we can reach the conclusion below:\\
1). The simulation result is in accordance with the theoretical value. \\
2). The simulation horizontal beam size hardly varies according to the polarity of SR.\\
3). The $\sigma$-polarization of SR should be utilized to measure the vertical beam size.\\
Thus, we select the $\sigma$-polarization of SR to measure the transverse beam size both for interferometer system and imaging system, both for horizontal size and vertical size.

\section{Results and analysis} \label{Results and analysis}
%文中给出了已经束流发生不稳定性时的尺寸与流强关系。还需补充束流稳定时，单束团模式束流尺寸与流强的关系。
%另外，应先介绍B7的成果。再介绍B8的结果。和前面的顺序保持一致。

\subsection{Measurement of beam size } \label{Measurement of beam size }
In the usual operation of HLS II, coupling correction is performed with 4 skew quadrupoles in order to obtain a beam lifetime of 10 hour at 300 mA. The coupling coefficient is about 10$\%$. Fig.\ref{fig4} and Fig.\ref{fig5} show fringes obtained by the vertical channel and the horizontal channel with a beam current of 100.4mA.

The raw data is a 2D matrix with 640 rows and 480 columns. And the data analysis is merely performed inside of a region of interest(ROI) area. Take vertical beam size analysis for instance. Set 11 columns data of the matrix which nearest to the peak position as the ROI area. With integrating the ROI along x direction, a vector which is similar with the blue curve as showed in Fig.\ref{fig4} can be obtained. By fitting the curve with Eq.~\eqref{eq1} and taking into account that imbalance between the intensities on the double slit, $\gamma$ can be inferred. The measured vertical beam size is $216\mu m(V)$(Fig.\ref{fig4}) and the horizontal beam size is 279 $\mu m$(Fig.\ref{fig5}).

The vertical beam size agrees with the theoretic value($212.0\mu m$) at 10$\%$ coupling coefficient. And the horizontal beam size is slightly larger than the theoretic value($246.9\mu m$). The reason might be the real energy spread is larger than the design value.

The possible error sources of the beam size measurement by the interferometer system contain the following aspect. Imperfection of the optic component and the finite pixel size of the CCD. Besides, the error of the distance between two slits(less than 1 $\mu m$), the slit size(less than 1 $\mu m$), the distance between the source potion and the double slit(10 mm), the nonlinearity of the CCD output(1.9$\%$ uncertainty) and the non-monochromatic light(2$\%$ uncertainty) will also result in measurement error. At last, beam jitter can generate a phase shift in the fringe and will reduce the visibility of the fringe\cite{14}. Assuming the beam offset from the orbit is 50 $\mu m$, we can get a visibility error of 1.7$\%$ and a horizontal beam size error of 1.4$\%$. The horizontal beam size measurement error due to above cause is 3.0$\%$.

\begin{center}
\includegraphics[width=8cm]{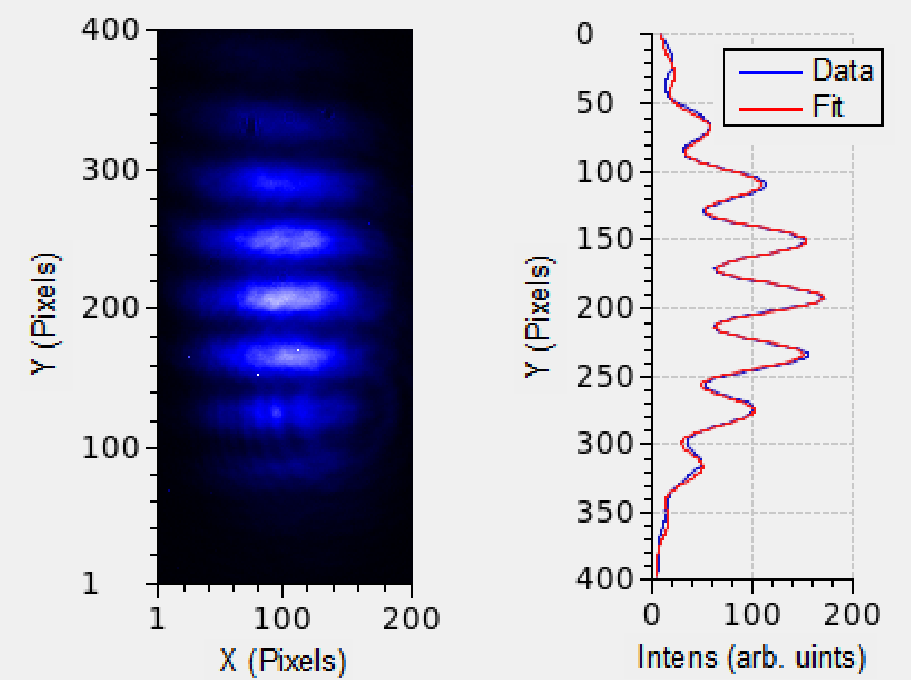}
\figcaption{\label{fig4}   A typical interference fringe obtained by the vertical interferometer using $\sigma$-polarized SR.}
\end{center}

\begin{center}
\includegraphics[width=8cm]{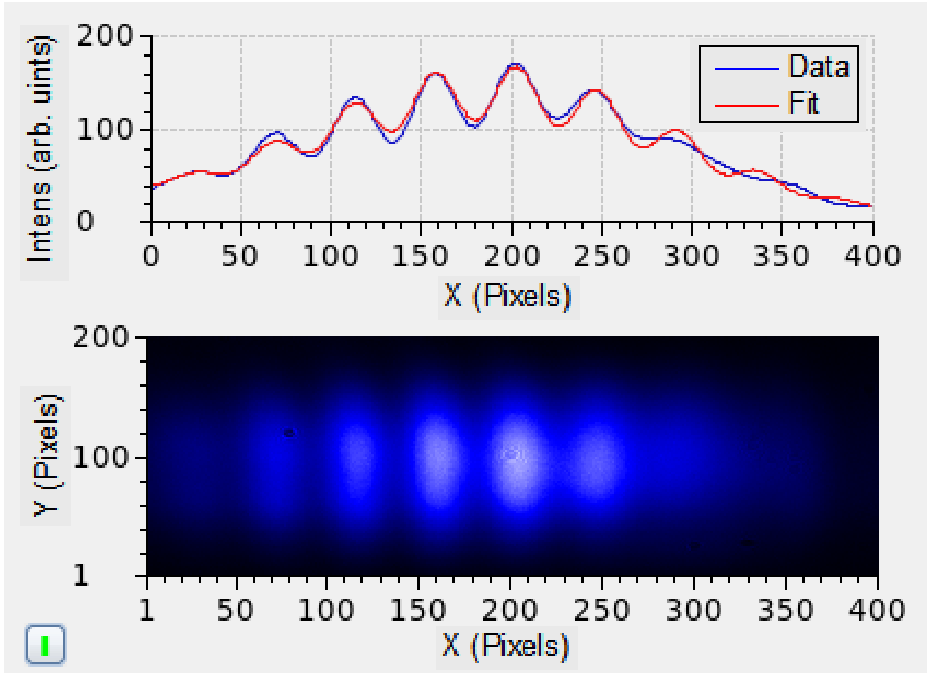}
\figcaption{\label{fig5}   A typical interference fringe obtained by the horizontal interferometer using $\sigma$-polarized SR.}
\end{center}

Beam sizes of B8 source point were measured at single bunch operation. A set of background images was acquired at different beam current range. Exposure time should change in case of saturation or underexposure. A typical beam size measurement is $303\mu m(H) \times 174\mu m(V)$(Fig.\ref{fig6}) with a beam current of 4.3 mA. Measurement of beam size at different beam current was obtained(Fig.\ref{fig7}). 100 images were acquired at each beam current. Measurement were repeated 3 times.

The vertical beam size(about 174$\mu m$) agrees with the theoretic value(159.5$\mu m$) and does not depend on the beam current. But the horizontal beam size is far larger than the theoretic value and depends on the beam current. It was found in the measurement that the beam became unstable when the beam current was more than 8 mA.

\begin{center}
\includegraphics[width=8cm]{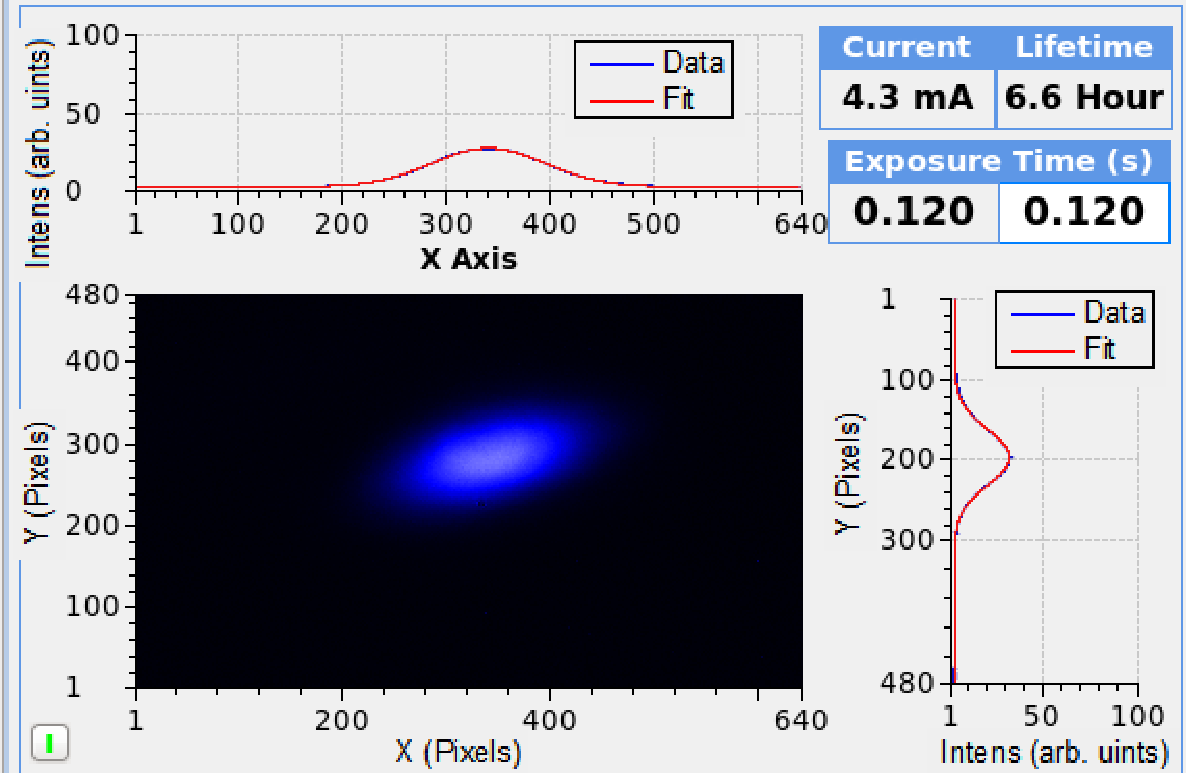}
\figcaption{\label{fig6} The measured intensity distribution of $\sigma$-polarized SR at the image plane and the best fit. The beam current is 4.3 mA(at single bunch operation).}
\end{center}

\begin{center}
\includegraphics[width=8cm]{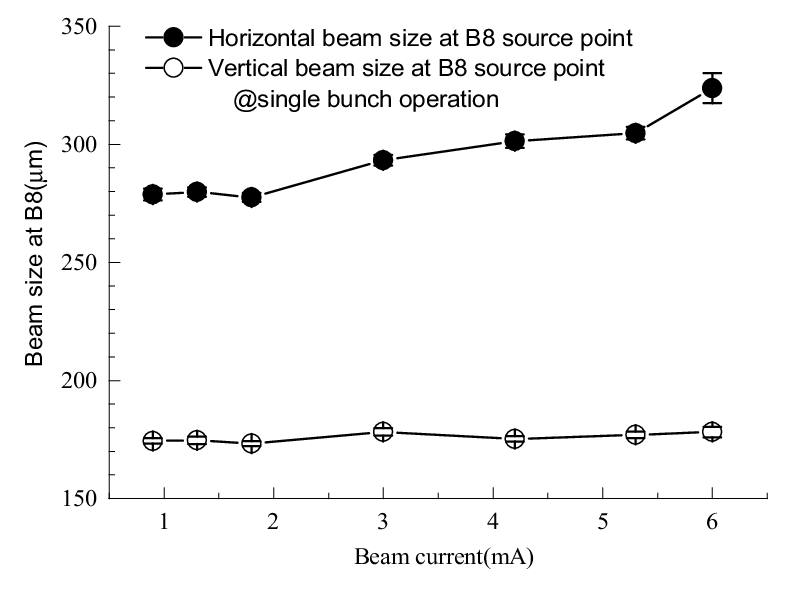}
\figcaption{\label{fig7}   The horizontal beam size and vertical beam size ($\pm$1 standard deviation) measured by the imaging system.}
\end{center}

\subsection{Measuring beam energy spread and beam emittance} \label{Measuring beam energy spread and beam emittance}
Horizontal beam sizes of the B7 and B8 source point were measured simultaneously at different beam current(from 80 mA to 240 mA with a step of 20 mA) at multi bunch operation(Fig.~\ref{fig8}). The beam will be in the presence of longitudinal instability if the beam current is less than 80 mA. And the degree of coherence will be less than 0.2 if the beam current is larger than 240 mA.

The measured horizontal beam size of the B7 and B8 source point are far larger than the theoretic one. From Section\ref{Measuring beam energy spread and beam emittance}, the measured energy spread is larger than the theoretic value($4.7 \times 10^{-4}$). Besides, the horizontal beam size would be widened in the presence of a beam instability. From the performance of the feedback system of HLS II, we know that the transverse beam instability is reduced perfectly while the longitudinal beam instability is not reduced completely.

Statistics of the horizontal beam size of B8 source point(at 120 mA) show that the standard deviation is 7.5 $\mu m$ and the range(maximum - minimum) is 35$\mu m$. Similarly, statistics of B7 show that the standard deviation is 1.5 $\mu m$ and the range is 4$\mu m$. Thus, the result of B8 has a larger jitter than B7. The reason might be that the dispersion function of B8 is larger than B7.

\begin{center}
\includegraphics[width=8cm]{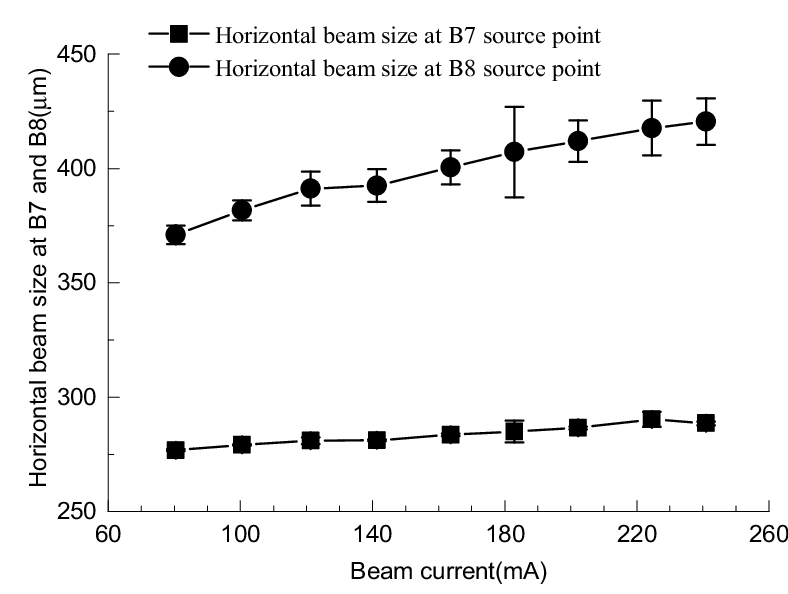}
\figcaption{\label{fig8}   The horizontal beam size($\pm$1 standard deviation) of the B7 source point measured by the interferometer and of the B8 source point measured by the imaging system (at multi bunch operation).}
\end{center}

The corresponding horizontal beam emittance and the beam energy spread are inferred(Fig.~\ref{fig9}). The horizontal beam emittance at 220 mA is obviously extraordinary(36.24 nm.rad). The rest point give a mean value of 35.04 $\pm$ 0.24 nm.rad. The relative beam energy spread has a positive correlation with the beam current(from 80 mA to 240 mA).

%The vertical beam emittance can be calculated from the vertical beam size of these two systems respective. The vertical beam emittance was 4.56 $\pm$ 0.04 nm.rad measured by the interferometer and 4.02 $\pm$ 0.20 nm.rad by the imaging system. % Thus ,the transverse beam emittance was 39.76 nm.rad and 39.22 nm.rad.

In order to precisely measure the beam size when the beam current is more than 240 mA, a new double slit with a less separation will be utilized. And the longitudinal beam feedback system should be optimized to make sure that the beam is stable.

\begin{center}
\includegraphics[width=8cm]{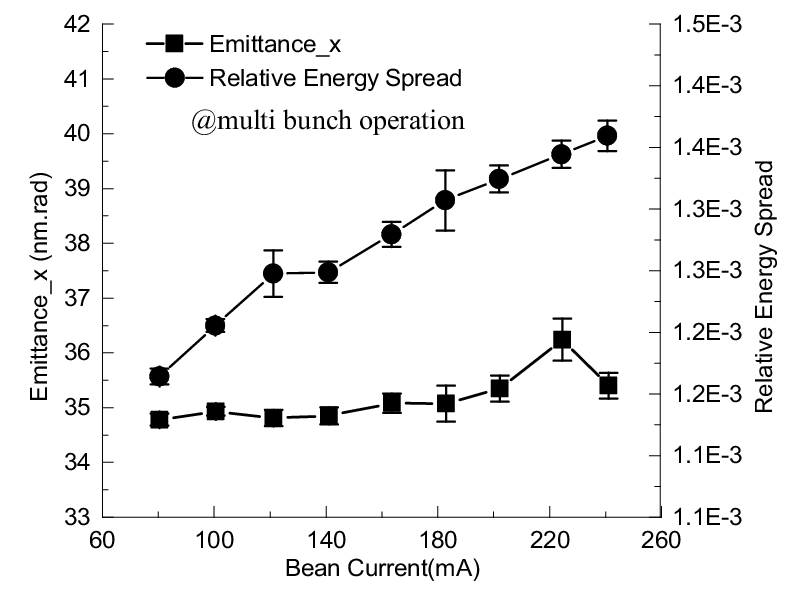}
\figcaption{\label{fig9} Horizontal emittance and beam relative energy spread($\pm$1 standard deviation) measured by the interferometer system and the imaging system(at multi bunch operation).}
\end{center}
%实际上，不同纵向位置处的耦合系数会有所不同。
\section{Conclusion} \label{Conclusion}
Two transverse beam profile measurement systems using visible SR have been installed at HLS II storage ring. One is a SR interferometer system consisted of two channels. Another is a simple imaging system which can monitor the beam transverse profile directly. Oxygen-free copper mirrors is applied to prevent thermal deformations from the SR heat load.

Simulations of these two optic systems were finished using SRW. The simulation results are consistent with the theoretic value. And the $\sigma$-polarized component of SR can be used in measurement.

Measurement of beam size at different beam current at single bunch operation are obtained. The vertical beam size agrees with the theoretic value at 10$\%$ coupling coefficient and the horizontal beam size is larger than the theoretic value. With combining the interferometer system and imaging system, the horizontal beam emittance and the beam energy spread are inferred.

\end{multicols}
\bibliography{20151228}

\begin{thebibliography}{10}

\bibitem{1}
G~Kube.
\newblock {Review of Synchrotron Radiation based Diagnostics for Transverse
  Profile Measurements}.
\newblock In {\em Proceedings of DIPAC 2007}, pages 6--10, Venice, Italy, 2007.

\bibitem{2}
S~Takano.
\newblock {Beam diagnostics with synchrotron radiation in light sources}.
\newblock In {\em Proceedings of IPAC 2010}, pages 2392--2396, Kyoto, Japan,
  2010.

\bibitem{3}
T~Naito and T~Mitsuhashi.
\newblock {Improvement of the Resolution of SR Interferometer at KEK-ATF
  Damping Ring}.
\newblock In {\em Proceedings of IPAC 2010}, volume~6, pages 972--974, Kyoto,
  Japan, 2010.

\bibitem{4}
L~L Tang.
\newblock {\em Development and Study of Beam Profile Measurement System for HLS
  II}.
\newblock PhD thesis, University of Science and Technology of China, 2013.

\bibitem{5}
T~Mitsuhashi.
\newblock Spatial coherency of the synchrotron radiation at the visible light
  region and its application for the electron beam profile measurement.
\newblock In {\em Proceeding of PAC of 1997}, volume~1, pages 766--768. IEEE,
  1997.

\bibitem{6}
T~Mitsuhashi and T~Naito.
\newblock {Measurement of beam size at the ATF damping ring with the SR
  interferometer}.
\newblock In {\em Proceeding of sixth EPAC}, volume~22, pages 1565--1567,
  Stockholm, Sweden, 1998.

\bibitem{7}
T~Mitsuhashi.
\newblock {Measurement of Small Transverse Beam Size Using Interferometry}.
\newblock In {\em Proceedings DIPAC 2001}, ESRF, Grenoble, France, 2001.

\bibitem{8}
T~Mitsuhashi.
\newblock {Twelve Years of SR Monitor Development at KEK}.
\newblock In {\em Proceedings of BIW 2004}, Oak Ridge, Tennessee, 2004.

\bibitem{9}
P~Chevtsov, A~Day, J~Denard, et~al.
\newblock {Non-invasive energy spread monitoring for the JLAB experimental
  program via synchrotron light interferometers}.
\newblock {\em Nuclear Instruments and Methods in Physics Research Section A},
  557(1):324--327, 2006.

\bibitem{10}
J~Corbett, W~Cheng, A~Fisher, et~al.
\newblock {Interferometer Beam Size Measurements in SPEAR3}.
\newblock In {\em Proceedings PAC 2009}, Vancouver, BC, Canada, 2009.

\bibitem{11}
S~T Wang, D~L Rubin, J~Conway, et~al.
\newblock {Visible-light beam size monitors using synchrotron radiation at
  CESR}.
\newblock {\em Nuclear Instruments and Methods in Physics Research Section A:
  Accelerators, Spectrometers, Detectors and Associated Equipment}, 703:80--90,
  2013.

\bibitem{12}
J~Chen, K~R Ye, and Y~B Leng.
\newblock {Development of Shanghai Synchrotron Radiation Facility synchrotron
  radiation interferometer}.
\newblock {\em HIGH POWER LASER AND PARTICLE BEAMS}, 23(1):179--184, 2011.

\bibitem{13}
L~L Tang, B~G Sun, Y~Y Xiao, et~al.
\newblock {The measurement of electron beam transverse sizes by synchrotron
  radiation interferometry for HLS II}.
\newblock {\em Nuclear Science and Techniques}, 8.23(4):193--198, 2012.

\bibitem{14}
{\AA}~Andersson, M~B{\"o}ge, A~L{\"u}deke, V~Schlott, and A~Streun.
\newblock {Determination of a small vertical electron beam profile and
  emittance at the Swiss Light Source}.
\newblock {\em Nuclear Instruments and Methods in Physics Research Section A:
  Accelerators, Spectrometers, Detectors and Associated Equipment},
  591(3):437--446, 2008.

\bibitem{15}
A~Hansson, E~Wall{\'e}n, and {\AA}~Andersson.
\newblock {Transverse electron beam imaging system using visible synchrotron
  radiation at MAX III}.
\newblock {\em Nuclear Instruments and Methods in Physics Research Section A},
  671:94--102, 2012.

\bibitem{16}
J~Corbett, W~Cheng, A~Fisher, et~al.
\newblock {Interferometer Beam Size Measurements in SPEAR3}.
\newblock In {\em Proceedings PAC 2009}, Vancouver, BC, Canada, 2009.

\bibitem{17}
L~Wang, J~X Zhao, J~S Cao, et~al.
\newblock {Beam size measurement of BEPC II storage ring by using visible
  synchrotron light interferometry}.
\newblock {\em HIGH POWER LASER AND PARTICLE BEAMS}, 23(9):2512--2514, 2011.

\bibitem{18}
M~Masaki and S~Takano.
\newblock {Two-dimensional visible synchrotron light interferometry for
  transverse beam-profile measurement at the SPring-8 storage ring}.
\newblock {\em Journal of synchrotron radiation}, 10(4):295302, 2003.

\bibitem{19}
T~Naito and T~Mitsuhashi.
\newblock Very small beam-size measurement by a reflective synchrotron
  radiation interferometer.
\newblock {\em Physical Review Special Topics-Accelerators and Beams},
  9(12):122802, 2006.

\bibitem{20}
K~Tang, J~G Wang, B~G Sun, et~al.
\newblock {Beam transverse size and emittance measurement of HLS II using
  interferometry}.
\newblock {\em HIGH POWER LASER AND PARTICLE BEAMS}, 27:245--249, 2014.

\bibitem{21}
O~Chubar, P~Elleaume, S~Kuznetsov, et~al.
\newblock {Physical optics computer code optimized for synchrotron radiation}.
\newblock In {\em International Symposium on Optical Science and Technology},
  pages 145--151. International Society for Optics and Photonics, 2002.

\end{thebibliography}

\clearpage
\end{CJK*}
\end{document}